\documentclass[10pt,conference]{IEEEtran}
\IEEEoverridecommandlockouts

\usepackage{xspace}
\usepackage{pifont}
\usepackage{booktabs,makecell,tabularx}
\usepackage{multirow}
\usepackage{graphicx}
\usepackage{enumitem}
\usepackage{fontawesome}
\usepackage{balance}
\usepackage{amsmath} 
\usepackage[hidelinks]{hyperref}

\newcommand{\appname}{{\sc Refploit}\xspace}
\newcommand{\appnamebold}{{\sc \textbf{Refploit}}\xspace}
\newcommand{\baselinename}{{\sc PoCGen}\xspace}
\newcommand{\baselinenamebold}{{\sc \textbf{PoCGen}}\xspace}

\begin{document}

\title{Refploit: Facilitating Exploit Construction via Code-Agent Trajectory Repair}

\author{
\IEEEauthorblockN{
Zirui Chen\IEEEauthorrefmark{1},
Zhipeng Xue\IEEEauthorrefmark{1},
Jiayuan Zhou\IEEEauthorrefmark{2},
Xing Hu\IEEEauthorrefmark{1}\textsuperscript{\ddag},
Xin Xia\IEEEauthorrefmark{1},
and Xiaohu Yang\IEEEauthorrefmark{1}
}
\IEEEauthorblockA{
\IEEEauthorrefmark{1}The State Key Laboratory of Blockchain and Data Security, Zhejiang University, Hangzhou, China\\
\{chenzirui, zhipengxue, xinghu, yangxh\}@zju.edu.cn, xin.xia@acm.org
}
\IEEEauthorblockA{
\IEEEauthorrefmark{2}Queen's University, Kingston, Canada\\
jiayuan.zhou@queensu.ca
}
\thanks{\textsuperscript{\ddag} Corresponding Author}
}

\maketitle

\begin{abstract}
Vulnerability exploits play a crucial role in assessing the downstream impact of Java library vulnerabilities.
While some vulnerabilities are accompanied by disclosed exploit references, automatically reproducing such references into runnable exploits remains challenging because they are often incomplete, unstructured, or only describe partial reproduction steps. Recent code agents provide a promising way to automate this process, but our study shows that their generated exploits often appear successful without triggering the actual vulnerable logic, such as replacing vulnerable APIs with self-implemented functions. To address this, we propose \appnamebold, an LLM-based trajectory recovery framework for facilitating vulnerability reproduction from public exploit references. The key insight is that a failed agent trajectory is not entirely useless. It may have already completed some reproduction subtasks while also revealing misleading directions that should be avoided. \appnamebold first validates an agent-generated exploit through differential execution. When the exploit is ineffective, \appnamebold analyzes its reproduction progress, locates the trajectory segments associated with the reproduction progress, and derives constraints to guide focused recovery. We evaluate \appnamebold on three open-source Java vulnerability datasets, covering 172 exploit references for 143 vulnerabilities. Under DeepSeek-V4-Flash, \appnamebold successfully reproduces 138 exploits, achieving a reproduction rate of 80.2\%. It achieves a 64.3\% relative improvement over the initially generated trajectories and outperforms both the SOTA exploit-generation method \baselinenamebold and advanced code agents such as Codex with GPT-5.4. Our ablation study shows that differential execution reduces false reproductions, while progress-guided trajectory analysis and constraint-guided recovery improve repair effectiveness and efficiency. We further adapt \appnamebold to another code agent and observe consistent improvements, demonstrating its generality.

\end{abstract}


\section{Introduction}

With the widespread adoption of open-source libraries~\cite{Synopsys1,Na2024component,kula2018developers, Zhang2024SymBisect,Zhang2025mitigation,cassel2025nodemedic,Zimmermann2019NPM,Wu2024Library}, the exploitability of upstream vulnerabilities in downstream projects has drawn increasing attention~\cite{ zirui2024exploiting,He2023Dependent,shuhan2025vul,Bavota1,kula2018developers,Li2023downstream,Nusrat2022npm}. This issue is particularly important in the Java ecosystem, where Maven~\cite{Maven} hosts ten million packages, making downstream impact assessment a critical task. Recent research employs exploits as evidence for assessing the impact of upstream vulnerabilities~\cite{Avgerinos1}, such as determining affected library versions~\cite{Dai2021Exploit,Jiang2023AEM,zirui2025poc} and providing domain knowledge for generating exploit tests in downstream projects~\cite{Zhou2024Magneto,zirui2024exploiting,Hong2022Poc, Deng2025Chainfuzz}.

However, manually reproducing vulnerabilities is often time-consuming and error-prone~\cite{Bhuiyan2023Exploit}. Existing studies mainly attempt to generate exploits based on vulnerability descriptions by designing validation strategies for specific vulnerability types~\cite{Michael2025PoC, Ray2025PoC}. However, such strategies have limited generality because vulnerability behaviors vary substantially across vulnerabilities and libraries, making it difficult to design general validation oracles. For example, \baselinename~\cite{Michael2025PoC} introduces five static analysis queries and reproduction patterns for path traversal (CWE-22/35), prototype pollution (CWE-1321), command injection (CWE-77/78), code injection (CWE-94 to CWE-99), and ReDoS (CWE-1333). This design limits the generality of \baselinename when the vulnerability requires specific validation beyond these predefined patterns.

To support reproduction across a broader range of vulnerability types, we examine reusable exploit knowledge beyond predefined patterns and find that public exploit references are available for some Java vulnerabilities. Specifically, among the 151 vulnerabilities in \textit{CWE-Bench-Java}~\cite{li2025iris} and \textit{VISION}~\cite{Wu2024Vision}, 72 vulnerabilities include exploit links in NVD. However, our empirical study shows that \textbf{these disclosed exploits are rarely directly executable}. They are often incomplete or unstructured, providing only textual reproduction steps, partial code snippets, or advisory-level descriptions. As a result, additional effort is required to reproduce the expected behavior, including completing the project, locating the vulnerable API, preparing the environment, and designing verification logic. On average, each reproduction requires 2.15 manual interventions, highlighting the need for automation.

Code agents have demonstrated promising capabilities in automatically building software repositories~\cite{Michael2025Setup, Yu2025Building}.
Motivated by this progress, we explore their potential for automating vulnerability reproduction from exploit reference. However, we find that commonly used code agents (such as mini-swe-agent) often produce outputs that appear to reproduce the vulnerability but do not exercise the actual required logic. For instance, when reproducing CVE-2018-1002200, the agent utilizes its own implementation of the ZIP extraction logic instead of invoking the vulnerable API, and for CVE-2021-39144, it triggers the vulnerability by adjusting configurations rather than using the intended payload. These observations motivate approaches that repair ineffective agent-generated trajectories to facilitate vulnerability reproduction.

By inspecting these ineffective agent trajectories, we find that \textbf{a failed reproduction attempt is not entirely useless}. The trajectory may have already completed some tasks required for reproduction, while exposing misleading directions that should be avoided in subsequent attempts. These signals can be distilled into preservation and repair constraints for trajectory refinement. Motivated by this insight, we propose \appname, an LLM-based trajectory recovery framework for facilitating vulnerability reproduction. \ding{182} \appname first executes the agent-generated exploit on both the selected and patched versions to determine whether the trajectory produces the expected differential behavior. \ding{183} When the trajectory is ineffective, \appname analyzes its reproduction progress and maps the resulting judgments back to trajectory segments. \ding{184} Based on this analysis, \appname derives constraints from the failed trajectory to guide recovery, specifying which directions should be preserved and which actions should be repaired.
Guided by these constraints, \appname resumes the agent with focused recovery tasks and iteratively re-validates the repaired exploit until it achieves effective reproduction.

We evaluate \appname on three open-source datasets with 172 exploits for 143 vulnerabilities. These vulnerabilities span 53 CWE types, covering at least one CWE associated with 61.9\% of Maven vulnerabilities in GitHub Advisory. Overall, \appname successfully reproduces 138 exploits, achieving a success rate of 80.2\%. It outperforms all baselines, including \baselinename on its predefined CWE types, the base agent mini-swe-agent (48.8\%), and Codex with GPT-5.4 (69.2\%). Our ablation results further show that each component contributes to the performance of \appname. Differential analysis reduces false reproductions, while progress assessment and constraint guidance reduce unnecessary repair attempts.
Finally, we evaluate the generality of \appname by adapting \appname to another code agent, which shows that \appname can consistently improve the initial trajectories across agents.

The main contributions of this paper are as follows.

\begin{itemize}[leftmargin=*]
\item We conduct a systematic study on the gap between disclosed exploit references and runnable exploits, showing that reproduction requires additional effort in environment preparation, harness assembly, and exploit behavior adaptation.

\item We propose \appname, a trajectory recovery framework for facilitating vulnerability reproduction by analyzing initial trajectories and deriving constraints to guide recovery.

\item On three vulnerability datasets with 172 exploits, \appname achieves a reproduction rate of 80.2\%, outperforming both code-agent baselines and the exploit-generation baseline.

\end{itemize}

\section{Preliminary Study}

In this section, we introduce the usage scenario, present our empirical study, and discuss a motivating example.

\subsection{Usage Scenario}

In real-world vulnerability management, our method is designed to support two practical usage scenarios.

\begin{figure*}[htbp] 
  \centering	
  
  \includegraphics[width=0.98\linewidth]{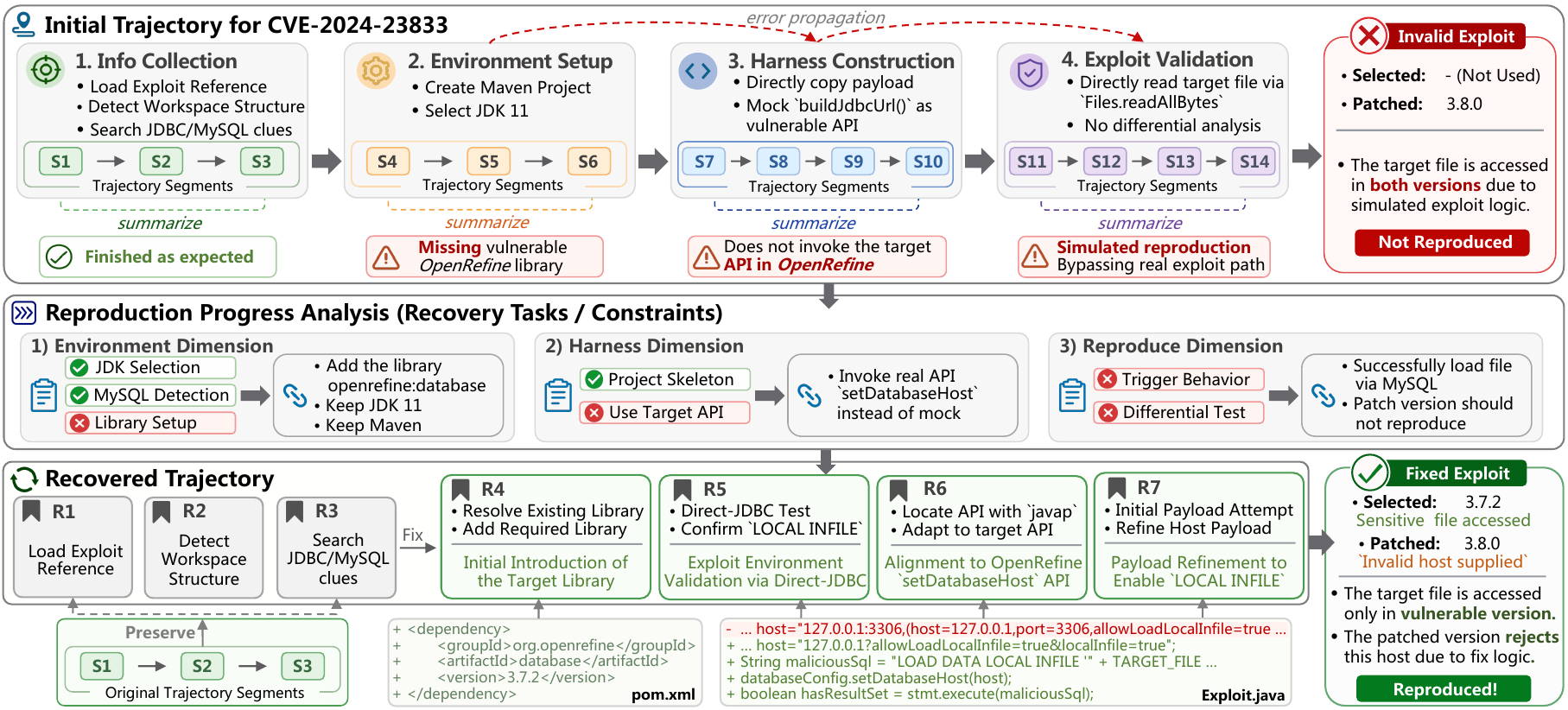}
  
  \caption{Refining the initial trajectory of reproducing CVE-2024-23833 by \appname.}
  \label{fig:example}
\end{figure*}

(1) Constructing runnable reproductions from incomplete disclosed exploit references. Public exploit references are important for validating whether a reported library vulnerability can be reproduced in practice~\cite{Dai2021Exploit,zirui2024exploiting,Zhou2024Magneto,zirui2025poc}, but they may contain only a textual description or a partial test snippet rather than a complete runnable project. In such cases, security analysts still need to assemble the harness, prepare the environment, and make the vulnerability observable. \appname supports this process by guiding a code agent to construct a valid reproduction attempt through differential execution.

(2) Refining ineffective vulnerability reproduction trajectories. When a code agent fails to generate a valid exploit, its trajectory is not entirely useless and may still contain segments that support vulnerability reproduction. \appname analyzes such trajectories to identify what can be reused, what remains missing, and what may mislead reproduction.
These signals guide constrained trajectory repair rather than restarting the reproduction process, thereby improving efficiency.

\subsection{Reproduction Process Study}

To understand the process of reproducing exploit references, we conduct a manual reproduction study on NVD exploit references associated with vulnerabilities from two Java vulnerability datasets, \textit{CWE-Bench-Java}~\cite{li2025iris} and \textit{VISION}~\cite{Wu2024Vision}, which together contain 151 unique vulnerabilities. After filtering inaccessible and irrelevant links, we obtain 84 exploit references. Two researchers with over five years of experience in vulnerability reproduction then attempt to convert each reference into a runnable exploit project, with a time limit of two hours per case. During this process, the researchers record the encountered issues and the remediation steps. To ensure reliability, the two researchers cross-check both the constructed exploits and the recorded interventions, and resolve disagreements through discussion. This process involves 159 interventions and results in 74 valid exploits.

\begin{table}[t]
\centering
\caption{Manual Interventions during Each Stage.}
\label{tab:manual_efforts}
\resizebox{0.95\linewidth}{!}{
\begin{tabular}{clcc}
\toprule
\textbf{Source Form} & \textbf{Intervention Category} & \textbf{Stage} & \textbf{Count} \\
\midrule

% Block 1: Textual Steps
% 注意：\shortstack[l] 表示左对齐，\\ 用于换行
\multirow{4}{*}{\shortstack[l]{\textbf{Textual Steps} \\ (14/17 success)}} 
 & Exploit Project Setup & Harness & 14 \\
 & Exploit Harness Construction & Harness & 9 \\
 & Exploit Resource Setup & Environment & 7 \\
 & Exploit Script Modification & Reproduce & 6 \\
\midrule

% Block 2: Test Snippet
\multirow{4}{*}{\shortstack[l]{\textbf{Test Snippet} \\  (27/27 success)}} 
 & Exploit Project Setup & Harness & 27 \\
 & Test Snippet Completion & Harness & 17 \\
 & Test Environment Setup & Environment & 2 \\
 & Payload Modification & Reproduce & 4 \\
\midrule

% Block 3: Exploit Script
\multirow{3}{*}{\shortstack[l]{\textbf{Exploit Script} \\ (17/20 success)}} 
 & Exploit Project Setup & Harness & 6 \\
 & Exploit Harness Construction & Harness & 2 \\
 & External Tool Configuration & Environment & 7 \\
\midrule

% Block 4: Advisory
\multirow{2}{*}{\shortstack[l]{\textbf{Advisory} \\ (16/20 success)}} 
 & Exploit Project Setup & Harness & 16 \\
 & Exploit Harness Construction & Harness & 42 \\
\midrule

\multicolumn{3}{r}{\textbf{Total Interventions}} & \textbf{159} \\
\bottomrule
\end{tabular}
}
\end{table}

We use these interventions to characterize the reproduction process rather than only measuring manual cost. During manual reproduction, we observe that a valid exploit must run under a suitable environment, invoke the vulnerability-relevant library logic through a runnable harness, and produce observable behavior that confirms the vulnerability. To systematically derive these requirements, the two researchers perform an open card-sorting process over the recorded remediation steps. Each intervention is first coded according to the concrete requirement it addresses, and semantically similar interventions are then discussed and merged into intervention categories. Finally, as illustrated in Table~\ref{tab:manual_efforts}, these categories are mapped to three reproduction dimensions: \textbf{Environment Preparation}, \textbf{Harness Assembly}, and \textbf{Exploit Reproduction}.

\ding{182} \textbf{Environment Preparation} captures whether the runtime prerequisites required by the exploit are available. Some reproductions depend on specific operating systems, JDK versions, network settings, or external tools. Since these prerequisites are often underspecified in public references, a reproduction attempt may fail even when the exploit logic is correct.

\ding{183} \textbf{Harness Assembly} captures whether the fragmented exploit information has been transformed into a runnable harness that invokes the vulnerability-relevant library logic. The required actions include preparing the target project, locating the vulnerable API, constructing malicious inputs, and generating executable exploit code. This dimension is crucial because an exploit may appear runnable while bypassing the actual vulnerable library path.

\ding{184} \textbf{Exploit Reproduction} captures whether the exploit triggers observable vulnerability behavior in the prepared environment. Even when the project and harness are ready, payloads or verification logic may still require adaptation. For example, path-traversal payloads may need to match the local file-system layout, and GUI-based indicators may need to be replaced with effects that are easier to observe automatically.

These observations suggest that exploit reproduction should not be judged only by whether the final exploit runs successfully. Instead, it should be assessed according to whether the environment is prepared, the harness invokes the corresponding logic, and the exploit produces expected behavior.

\vspace{-0.6em}
\begin{center}
\resizebox{\linewidth}{!}{
\begin{tabular}{l!{\vrule width 1pt}p{0.9\columnwidth}}
\makecell{{\large \faLightbulbO}}  &\textbf{Findings:} Vulnerability reproduction follows a multi-dimensional process involving environment preparation, harness assembly, and exploit reproduction. These dimensions provide perspectives for assessing reproduction progress and diagnosing failed agent trajectories.\\
\end{tabular}}
\end{center}

\begin{figure*}[htbp] 
  \centering	
  
  \includegraphics[width=\linewidth]{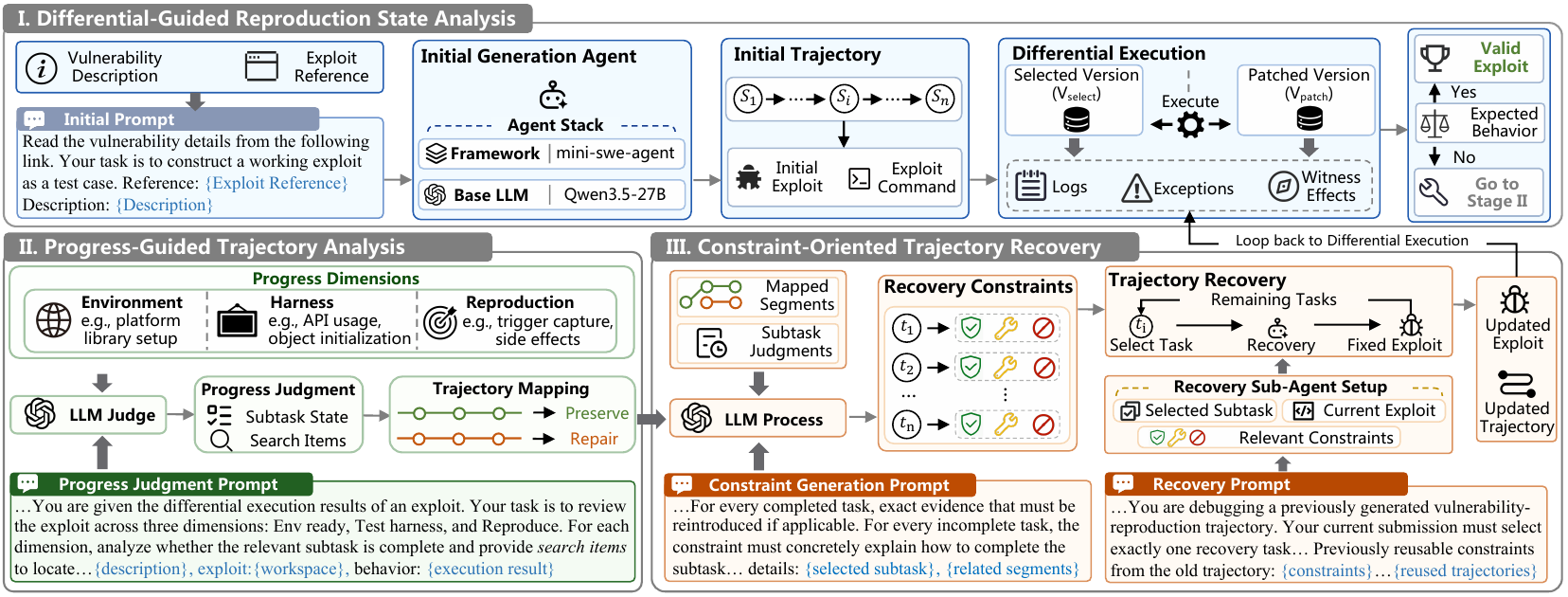}
  
  \caption{The overall framework and prompt details of \appname.}
  \label{fig:overview}
\end{figure*}

\subsection{Motivation Example}

As illustrated in Figure~\ref{fig:example}, we present an example where a code agent fails to reproduce a vulnerability even though the agent states that the generated exploit  ``\textit{clearly shows the vulnerability behavior}'' before finishing the task. CVE-2024-23833 is a JDBC attack vulnerability in OpenRefine, which arises because OpenRefine builds the MySQL JDBC URL by directly placing the user-provided host value into the URL without sufficient validation. For example, a host value containing \textit{(allowLoadLocalInfile=true)} can be interpreted by the driver as enabling local file loading, which may allow reading sensitive files. The vulnerability has been addressed since version 3.7.7. We find that the exploit does not exhibit differential behavior after the library is updated.

In the Initial Trajectory shown in Figure~\ref{fig:example}, we segment the trajectory into 14 segments based on two types of actions (reading the reference and running exploits). We then manually categorize each segment according to its purpose and examine whether it achieves the expected goal. We find that the initial trajectory has already achieved several goals, such as reading the exploit reference and constructing the Maven project skeleton. Meanwhile, we identify two reasons for the ineffective attempt. \ding{182} First, the trajectory does not follow the exploit path mentioned in the reference, which requires invoking the vulnerable library API \textit{setDatabaseHost} with a crafted MySQL host string. Instead, the generated project does not include the \textit{OpenRefine} library at all (Summary for \textbf{Environment Setup} in Figure~\ref{fig:example}). It directly implements a local mock connection manager whose \textit{buildJdbcUrl()} method manually inserts the user input into a JDBC URL to simulate the behavior of \textit{OpenRefine} (Summary for \textbf{Harness Construction} in Figure~\ref{fig:example}). \ding{183} Second, to demonstrate the expected malicious behavior, the harness directly reads a local file using standard Java file I/O with \textit{Files.readAllBytes(...)}, instead of triggering file access through the JDBC attack path described in the reference (Summary for \textbf{Exploit Validation} in Figure~\ref{fig:example}). 

This example shows that \ding{182} \textbf{the initial trajectory is not entirely useless, even though it fails to produce an effective differential result}. It already contains reusable components, such as selecting MySQL/JDBC as the attack direction, creating a Maven project, and targeting the correct vulnerable version. We further find that \ding{183} \textbf{the expected workflow of vulnerability reproduction provides useful dimensions for diagnosing trajectory progress}, which is crucial for determining whether each trajectory segment satisfies the requirements of vulnerability reproduction. By checking environment preparation, harness construction, and reproduction observation, we can determine which segments have achieved their intended goals, which segments introduce misleading behavior, and which reproduction requirements remain missing. Based on this analysis, we can extract refinement constraints that the next reproduction attempt should follow. These observations motivate our progress-guided trajectory analysis.

Based on these observations, we design \appname to \ding{182} generate constraints from reproduction progress analysis and \ding{183} recover the trajectory based on these constraints.  After extracting constraints from the initial trajectory, \appname first preserves reusable components from the initial trajectory. It then repairs the ineffective parts according to the constraints. For the environment dimension, \appname introduces the \textit{OpenRefine:3.7.2} library. For the harness dimension, \appname uses \textit{setDatabaseHost} to configure the crafted host. For the reproduction dimension, \appname constructs a payload based on the \textit{LOAD DATA LOCAL INFILE} command to load the target file, and modifies the payload to trigger sensitive file access. \appname confirms recovery by verifying that the repaired exploit reads the target file on the vulnerable version but is blocked by the patched version.

\section{Methodology}

As illustrated in Figure~\ref{fig:overview}, \appname employs a three-stage workflow to evaluate and repair vulnerability reproduction trajectories generated by code agents. First, it evaluates the initially generated exploit through differential execution on the selected and patched versions. Second, when the differential result is ineffective, \appname decomposes the trajectory into segments and aligns the reproduction process with trajectory segments to identify satisfied, missing, or misleading progress. Third, based on the completion status of each reproduction task in the original trajectory, \appname generates preservation and repair constraints and uses them to resume the agent with focused recovery guidance. The recovery loop iteratively applies constrained repairs and differential validation until the exploit reaches the expected reproduction behavior or the recovery budget is exhausted.

\subsection{Differential-Guided Reproduction State Analysis}

Given an initial reproduction trajectory generated by the code agent, this stage evaluates the trajectory through differential execution. \appname executes the generated exploit on both the target and patched versions, compares the observed behaviors with the expected vulnerability behavior, and collects differential evidence that indicates whether the trajectory reaches a successful reproduction state.

\subsubsection{Initial Trajectory Generation}

Given a vulnerability description $d$ and an exploit reference $r$, \appname first employs the code agent to generate an initial reproduction trajectory based on its ReAct loop:
\[
T_{\text{init}}=\langle S_1,S_2,\ldots,S_n\rangle.
\]
In each iteration, the agent generates the next environment command based on the vulnerability information $(d,r)$ and the previously generated trajectory prefix $\langle S_1,\ldots,S_{i-1}\rangle$, executes the command within the environment, and integrates the resulting observation into its subsequent decisions.

Each node $S_i$ records the intent for this step, the command generated by the agent, and the execution result. This information supports our subsequent trajectory analysis, constraint extraction, and trajectory recovery. As a result, $T_{\text{init}}$ captures both the generated initial exploit project $e_{\text{init}}$  and the step-by-step process by which the agent constructs the exploit.

\subsubsection{Differential Execution}

After generating $e_{\text{init}}$, \appname evaluates it on both the vulnerable version $V_{\text{vul}}$ used during its construction and the corresponding patched version $V_{\text{pat}}$:

\[
R_{\text{vul}} = \mathcal{E}(e_{\text{init}}, V_{\text{vul}}), \qquad
R_{\text{pat}} = \mathcal{E}(e_{\text{init}}, V_{\text{pat}})
\]

Here, $\mathcal{E}$ denotes the execution process, and $R_{\text{vul}}$ and $R_{\text{pat}}$ summarize the observable execution behaviors under the two versions, including execution logs, exceptions, and witness side effects. \appname then analyzes the differential behaviors to determine whether the observed execution matches the expected reproduction behavior mentioned in $d$.

To perform this assessment, the LLM receives the differential evidence $(R_{\text{vul}}, R_{\text{pat}})$ together with the vulnerability description $d$ and contextual information of $e_{\text{init}}$, including the project structure, generated exploit files, and the exploit command. The analysis focuses only on whether the exploit exhibits the intended behavior, while ignoring unrelated environmental, compilation, or execution issues. For cases that do not satisfy the expected behavior, \appname further assesses the reproduction progress to identify directions for trajectory refinement, repairs the trajectory accordingly, and re-validates the repaired exploit via differential execution.

\subsection{Progress-Guided Trajectory Analysis}

Given the differential evidence, \appname analyzes the current reproduction trajectory to extract evidence that can support reproduction and steer subsequent repairs away from repeated errors. We first define evidence-oriented progress dimensions that capture the necessary conditions for vulnerability reproduction, and then describe how \appname judges these dimensions and identifies reproduction-supporting and misleading trajectory components based on the evidence.

\subsubsection{Evidence-Oriented Progress Dimensions}

After differential execution, \appname does not immediately repair the ineffective trajectory. Instead, it first estimates how far the current attempt has progressed toward a valid reproduction. This distinction is important because an ineffective attempt does not mean that the whole trajectory is useless. Analyzing this progress allows \appname to classify the trajectory according to reproduction progress, identifying which components support reproduction, which requirements remain missing, and which components may mislead subsequent repair.

% 如何划分复现轨迹的阶段
Following prior work on subgoal-based agent evaluation~\cite{ma2024agentboard}, we decompose the vulnerability reproduction process into three progress dimensions based on the reproduction workflow summarized from our preliminary study:

\begin{itemize}[leftmargin=*]
    \item \textbf{Environment.} This dimension checks whether the reproduction conditions are available, including the Java runtime, library setup, and external services.

    \item \textbf{Harness.} This dimension checks whether the generated exploit forms a runnable harness that invokes the vulnerable library API and preserves the exploit semantics.

    \item \textbf{Reproduction.} This dimension checks whether the execution triggers the expected behavior and shows meaningful divergence between vulnerable and patched versions.
\end{itemize}

\subsubsection{Reproduction Progress Judgment}

To perform this assessment, \appname instructs the LLM to conduct a constrained reproduction progress judgment over the three dimensions. The analyzer reuses the same evidence context from differential execution, including $(R_{\text{vul}}, R_{\text{pat}})$, $d$, $r$, and the workspace context of $e_{\text{init}}$. \appname requires the LLM to expand each dimension $\delta_j$ into concrete reproduction subtasks based on $r$ and judge the progress of each subtask separately. For each dimension $\delta_j$, \appname derives a subtask set
$\mathcal{T}_j=\{t_{j,k}\}_{k=1}^{n_j}$, where each $t_{j,k}$ denotes a reproduction subtask under $\delta_j$.
For example, environment subtasks may involve preparing the correct Java runtime or external service, harness subtasks may involve invoking the vulnerable API as required, and reproduction subtasks may involve changing the payload to make the behavior observable.

For each $t_{j,k}$, \appname analyzes its completion status and the search items $\kappa_{j,k}$ used to localize the trajectory step that introduced the corresponding completed capability or issue. The completion status indicates whether the subtask has been satisfied, remains incomplete, or has been incorrectly completed. This judgment is directly usable for later trajectory repair. Subtasks marked as satisfied identify steps that can support reproduction and should be considered for preservation. Incomplete subtasks indicate missing requirements that should guide repair, while misleading subtasks identify directions that should be avoided in subsequent recovery.

\subsubsection{Progress-to-Trajectory Mapping}

Rather than being completed by a single command, a subtask may be supported by several related trajectory steps, such as inspecting the required environment information before editing the exploit harness. Therefore, \appname first decomposes the trajectory into segments based on two key actions: \ding{182} reading the initial vulnerability reference and \ding{183 }running the exploit commands. These actions indicate two important milestones in the trajectory: \ding{182} the agent is collecting exploit-relevant information and \ding{183} the agent considers the current exploit sufficiently complete for a reproduction attempt.

For each $t_{j,k}$, \appname uses its search items $\kappa_{j,k}$ to map the progress judgment back to the trajectory. If $\kappa_{j,k}$ is non-empty, \appname locates the earliest trajectory steps that introduce these items (i.e., the commands that write $\kappa_{j,k}$ into source files). Segments containing these steps are treated as the related trajectory evidence for $t_{j,k}$. If an incomplete subtask has no supporting evidence, \appname marks it as missing, indicating that the required capability has not been established in the current trajectory.

After mapping, each $t_{j,k}$ is associated with its completion status and related trajectory segments. In this way, \appname transforms the initial trajectory that is difficult to directly analyze into structured components categorized by their contribution to the reproduction process, which provides preservation, repair, and avoidance guidance that constrain the search space for recovery.

\subsection{Constraint-Oriented Trajectory Recovery}

After progress-guided trajectory analysis, \appname uses the mapped trajectory segments and judgments to drive recovery. This stage derives recovery constraints, selects a replay entry point, iteratively chooses repair tasks, and resumes the trajectory under the corresponding constraints.

\subsubsection{Recovery Constraint Generation}

Based on the mapped progress judgments, \appname converts trajectory segments into actionable recovery constraints that provide explicit instructions for the resumed agent during recovery.

\ding{182} For each completed subtask, \appname derives \textbf{preservation constraints} from its related trajectory segments. These constraints specify validated capabilities that should be retained instead of rediscovered. For example, if the initial trajectory has already selected a compatible JDK, constructed a Maven project skeleton, or identified MySQL/JDBC as the relevant attack direction, \appname instructs the LLM to distill these elements into preservation constraints, requiring the resumed agent to keep these capabilities when replaying the corresponding subtask during recovery.

\ding{183} For each incomplete subtask, \appname derives \textbf{repair constraints} that describe the missing requirement needed to advance reproduction. For example, it may require adding the real vulnerable library or adapting the payload so that the malicious behavior is triggered through the expected path. When the mapped trajectory evidence indicates that some segments drift away from the specified reproduction requirements, \textbf{repair constraints} further prevent the resumed agent from repeating misleading actions. For example, if the initial trajectory simulates vulnerable behavior through a locally implemented function, \appname forbids these directions in subsequent recovery during harness construction.

Together, these constraints transform the failed trajectories into recovery guidance. Preservation constraints encode validated progress, repair constraints encode missing requirements, and avoidance constraints encode failure-inducing directions that should not be repeated.

\subsubsection{Constraint-Guided Trajectory Recovery}

Given the generated recovery constraints, \appname first selects a replay entry point for recovery by scanning the initial trajectory from the beginning and replaying segments as long as they correspond to completed subtasks. Once \appname encounters a segment associated with incomplete or incorrectly completed subtasks, it resumes recovery from that point. Otherwise, when no reliable completed prefix can be identified, \appname starts a new recovery trajectory from the initial workspace state to avoid inheriting misleading trajectory content.

During recovery, \appname organizes the reproduction subtasks into a constraint-guided recovery plan, where each subtask is associated with the constraints derived from the trajectory analysis. In each recovery iteration, \appname selects one $t_{j,k}$ as the current repair target and launches a recovery sub-agent to work on this target under the associated constraints. The sub-agent receives the vulnerability context, the current workspace, and the constraints relevant to $t_{j,k}$. It is required to focus on $t_{j,k}$, preserve validated capabilities, and avoid repeating misleading actions identified from the initial trajectory according to the constraints.

After the recovery sub-agent submits the repair for a selected $t_{j,k}$, \appname integrates the updated workspace and continues with the remaining subtasks. Once the current recovery round finishes, \appname returns to the differential execution stage and re-evaluates the repaired exploit on the selected and patched versions. If the differential result becomes effective, the recovery terminates successfully. Otherwise, \appname starts the next recovery round by performing progress-guided trajectory analysis again until the exploit is successfully reproduced or the recovery budget is exhausted.

\section{Experiment Setup}

This section presents the experimental design used to evaluate \appname. We first introduce the research questions (RQs) and then describe the dataset, implementation details, baselines, and ground-truth construction during evaluation.
Our evaluation aims to answer the following RQs:

\begin{itemize}[leftmargin=*]

\item {\textbf{RQ1 (Effectiveness):} How effective is \appname in recovering ineffective reproduction trajectories, and how does it compare with existing vulnerability reproduction methods?}

\item {\textbf{RQ2 (Ablation Study):} How does each component contribute to the overall performance of \appname?}

\item {\textbf{RQ3 (Generality):} Can \appname achieve similar improvements when adapted to other code agents?}

\end{itemize}

\begin{table*}[t]
\centering
\caption{Effectiveness Comparison of \appname and Other Agents Across First-listed CWE Categories.}
\label{tab:effectiveness_cwe}
\small
\resizebox{\linewidth}{!}{
\begin{tabular}{cccccccccccccc}
\toprule
 \textbf{Method} & \textbf{Model}
& \makecell{\textbf{Overall}\\\textbf{\# Exploits=172}}
& \makecell{\textbf{CWE-22}\\(34)}
& \makecell{\textbf{CWE-787}\\(18)}
& \makecell{\textbf{CWE-502}\\(16)}
& \makecell{\textbf{CWE-79}\\(11)}
& \makecell{\textbf{CWE-611}\\(10)}
& \makecell{\textbf{CWE-770}\\(8)}
& \makecell{\textbf{CWE-94}\\(6)}
& \makecell{\textbf{CWE-78}\\(5)}
& \makecell{\textbf{CWE-444}\\(4)}
& \makecell{\textbf{CWE-776}\\(4)}
& \makecell{\textbf{Other}\\(56)}
 \\

\midrule

\multirow{2}{*}{\makecell[l]{\textbf{\appnamebold}}}  & \makecell[l]{DeepSeek-V4}
  & 138 (80.2\%) & 25 (73.5\%) & 15 (83.3\%) & 13 (81.2\%) & 10 (90.9\%) & 10 (100.0\%) & 8 (100.0\%) & 4 (66.7\%) & 5
  (100.0\%) & 4 (100.0\%) & 4 (100.0\%) & 40 (71.4\%) \\

   & Qwen3.5-27B
    & 96 (55.8\%) & 22 (64.7\%) & 11 (61.1\%) & 8 (50.0\%) & 7 (63.6\%) & 10 (100.0\%) & 2 (25.0\%) & 1 (16.7\%) & 3 (60.0\%) & 3 (75.0\%) & 3 (75.0\%) & 26 (46.4\%) \\

\midrule

 \multirow{2}{*}{\makecell[l]{\textbf{Codex (xhigh)}}} & \makecell[c]{GPT-5.4}
  & 119 (69.2\%) & 26 (76.5\%) & 12 (66.7\%) & 8 (50.0\%) & 8 (72.7\%) & 10 (100.0\%) & 6 (75.0\%) & 2 (33.3\%) & 0
  (0.0\%) & 4 (100.0\%) & 3 (75.0\%) & 40 (71.4\%) \\
 & \makecell[l]{DeepSeek-V4}
  & 90 (52.3\%) & 15 (44.1\%) & 12 (66.7\%) & 9 (56.2\%) & 3 (27.3\%) & 4 (40.0\%) & 4 (50.0\%) & 2 (33.3\%) & 4 (80.0\%) & 2 (50.0\%) & 3 (75.0\%) & 32 (57.1\%) \\

\midrule

 \multirow{2}{*}{\makecell[l]{\textbf{mini-swe-agent}}}  & DeepSeek-V4
  & 84 (48.8\%) & 20 (58.8\%) & 8 (44.4\%) & 10 (62.5\%) & 2 (18.2\%) & 4 (40.0\%) & 6 (75.0\%) & 2 (33.3\%) & 2 (40.0\%) & 2 (50.0\%) & 3 (75.0\%) & 25 (44.6\%) \\
 & Qwen3.5-27B
  & 47 (27.3\%) & 9 (26.5\%) & 5 (27.8\%) & 3 (18.8\%) & 6 (54.5\%) & 2 (20.0\%) & 2 (25.0\%) & 1 (16.7\%) & 0 (0.0\%) & 2 (50.0\%) & 2 (50.0\%) & 15 (26.8\%) \\

\midrule

\multirow{2}{*}{\makecell[l]{\textbf{AutoGPT}}}  & DeepSeek-V4
  & 48 (27.9\%) & 11 (32.4\%) & 9 (50.0\%) & 5 (31.2\%) & 4 (36.4\%) & 1 (10.0\%) & 1 (12.5\%) & 2 (33.3\%) & 0 (0.0\%) & 1 (25.0\%) & 1 (25.0\%) & 13 (23.2\%) \\
  & Qwen3.5-27B
  & 39 (22.7\%) & 4 (11.8\%) & 8 (44.4\%) & 2 (12.5\%) & 1 (9.1\%) & 3 (30.0\%) & 1 (12.5\%) & 0 (0.0\%) & 0 (0.0\%) & 2 (50.0\%) & 2 (50.0\%) & 16 (28.6\%) \\

\bottomrule
\end{tabular}}
\end{table*}

\subsection{Dataset}

To ensure sufficient dataset coverage, we conduct experiments on three open-source Java library vulnerability datasets~\cite{li2025iris, Wu2024Vision, chen2026largescaleempiricalstudygeneralizability}. Together, these three datasets cover a total of 278 CVEs. We collect all exploit links disclosed on NVD for these CVEs, obtaining 199 links for 160 CVEs. After excluding inaccessible links, links unrelated to exploit reproduction, links whose key information is available only in images, and two CVEs whose patches do not directly eliminate the vulnerable behavior (CVE-2021-43859 and CVE-2022-22885, as reported by Chen et al.~\cite{chen2026largescaleempiricalstudygeneralizability}), we obtain 172 exploit reference links corresponding to 143 CVEs. These CVEs cover 53 CWE types, which cover at least one CWE for 61.9\% of Maven library vulnerabilities in GitHub Advisory. Therefore, we consider this dataset sufficiently general for evaluation.

\subsection{Implementation}

In RQ1, we evaluate \appname with two open-weight models to support reproducible comparison. \ding{182} \textit{DeepSeek V4-Flash}~\cite{xu2026deepseek} is selected as a strong open-weight model with competitive coding and reasoning capabilities. \ding{183} \textit{Qwen3.5-27B}~\cite{qwen3.5} is selected because it can be locally deployed for reproducible evaluation, while also providing strong tool-calling capability and competitive performance on coding tasks. For RQ2 and RQ3, we use \textit{Qwen3.5-27B} as the default model setting, because local deployment enables stable and cost-controllable evaluation. We deploy Qwen3.5-27B on a machine with eight A100 GPUs, while accessing the remaining models through APIs. We set the temperature to 0.2 for all models.  For settings that require other models, such as Codex, we provide the details in the baseline description.

For our choice of code agent, we adopt \textbf{mini-swe-agent} (v2.2.8)~\cite{yang2024sweagent} as the base agent, as it is a lightweight and efficient code agent that facilitates reproducibility and supports multiple foundational models. For cross-agent generalizability evaluation, we include a widely used open-source code agent, \textbf{OpenHands}~\cite{wang2025openhandsopenplatformai} (74.5K stars on GitHub).

For each reproduction, we launch an Ubuntu 22.04 container with four LTS versions of Java. In all experiments, we provide only the vulnerability description and the exploit reference without supplying additional guidance.
Following prior work~\cite{Michael2025PoC}, we allocate a token budget of 5M input tokens and 1M output tokens, which corresponds to an upper-bound cost of approximately \$1.03 per case (\textit{DeepSeek-V4-Flash}), excluding cache hits. To prevent unbounded exploration, we limit the agent execution trajectory to at most 200 steps.

\subsection{Baselines}

To evaluate the performance of \appname in vulnerability reproduction, we consider the following baselines under the same budget and runtime environment:

\begin{itemize}[leftmargin=*]

\item \textbf{Exploit Generation Baselines.}
We include \baselinename~\cite{Michael2025PoC} because it represents the SOTA approach for automated exploit generation. We adapt \baselinename to Java and evaluate \baselinename under \textit{DeepSeek-V4-Flash} for a fair comparison.

\item \textbf{Advanced Code-Agent Baseline.}
To compare \appname with an advanced code-agent setting, we evaluate \textbf{Codex (0.141.0)}~\cite{codex} under the xhigh-context setting using its recommended model \textit{GPT-5.4} and the model used by \appname, \textit{DeepSeek V4-Flash} to ensure a fair comparison.

\item \textbf{Underlying code-agent baseline.}
To quantify the improvement of \appname over its underlying code agent, we directly evaluate \textbf{mini-swe-agent} under \textit{DeepSeek V4-Flash} and \textit{Qwen3.5-27B}, which measures the performance of the base agent without trajectory recovery.

\item \textbf{Open-source Code-Agent Baselines.}
 Following Simsek et al.~\cite{Michael2025PoC}, we select \textbf{AutoGPT Classic}~\cite{AutoGPT} as a code-agent baseline and evaluate it under \textit{DeepSeek V4-Flash} and \textit{Qwen3.5-27B}, which can navigate the codebase by browsing files, reading and writing files, and executing commands.

\end{itemize}

\subsection{Ground Truth}

Due to the wide variety of vulnerability types in our dataset, it is impractical to design vulnerability-specific validation rules for every case. Existing strategies, such as rule-based validation used by Simsek et al.~\cite{Michael2025PoC}, cover only five CWEs. To evaluate exploit reproduction, we adopt a differential-based strategy. Specifically, we execute each exploit on both the vulnerable version and the patched version. After collecting the execution behaviors, two researchers with over five years of experience in software security independently analyze the results. To mitigate potential bias, we adopt a blinded annotation strategy. For each reproduction task, all generated exploits from different settings, including \appname, baselines, ablation variants, and cross-agent variants, are randomly ordered and presented to the annotators without revealing which configuration produced them. An exploit is considered reproduced if it triggers the intended behavior in the vulnerable version while failing to do so in the patched version. After the independent analyses, the two researchers cross-verify their results and resolve discrepancies through discussion. The two researchers achieve a Cohen's kappa of 0.828, indicating strong agreement. After resolving discrepancies, we identify 1,114 confirmed reproductions in total.

\section{Evaluation Results}

We evaluate the performance of \appname in terms of effectiveness, component contribution, and cross-agent generality.

\subsection{Effectiveness}

Table~\ref{tab:effectiveness_cwe} and Table~\ref{tab:CWE} present the effectiveness across CWE categories. Overall, \appname with \textit{DeepSeek-V4-Flash} achieves the best performance, reproducing 138 out of 172 exploits with a success rate of 80.2\%. It outperforms all evaluated baselines, including the exploit-generation baseline \baselinename on predefined categories, the widely used AutoGPT (27.9\%), the base agent mini-swe-agent with the same model setting (48.8\%), and Codex with GPT-5.4 (69.2\%). This result shows that \appname can achieve competitive effectiveness against advanced code agents and exploit reproduction methods.

To further control the effect of backbone models, we compare all code-agent baselines under \textit{DeepSeek-V4-Flash}. \appname achieves a success rate of 80.2\%, outperforming Codex, mini-swe-agent, and AutoGPT by 27.9\%, 31.4\%, and 52.3\%, respectively. This result shows that the improvement does not simply come from using a stronger backbone model.
The comparison with the underlying models further shows that \appname is effective across models. Under \textit{DeepSeek-V4-Flash}, REFPLOIT improves over the base agent from 84 to 138 reproductions, yielding 54 additional exploits and a 64.3\% relative increase. Under \textit{Qwen3.5-27B}, \appname improves reproductions from 47 to 96, a 104.3\% relative increase. These improvements show that \appname can effectively recover trajectories generated by different models.

\begin{table}[t]
\centering
\caption{Comparison with \baselinename Across Predefined CWE Types.}
\label{tab:CWE}
\small
\resizebox{\linewidth}{!}{
\begin{tabular}{ccccc}
\toprule
\textbf{Vulnerability Type}
& \textbf{CWE}
& \textbf{Exploits}
& \textbf{\appnamebold}
& \textbf{\baselinenamebold} \\

\midrule

Path Traversal
& CWE-22/35
& 35
& 26 (74.3\%)
& 11 (31.4\%) \\

Command Injection
& CWE-77/78
& 7
& 5 (71.4\%)
& 2 (28.6\%) \\

Code Injection
& CWE-94 to 99
& 10
& 8 (80.0\%)
& 3 (30.0\%) \\

ReDoS
& CWE-1333
& 3
& 3 (100.0\%)
& 2 (66.7\%) \\

Prototype Pollution
& CWE-1321
& N/A
& N/A
& N/A \\

\bottomrule
\end{tabular}}
\end{table}

Across CWE categories, \appname also demonstrates broad effectiveness compared with code-agent baselines. With \textit{DeepSeek-V4-Flash}, \appname achieves the best performance in most listed CWEs, outperforming most baselines across diverse vulnerability categories such as deserialization, XXE, resource management, and HTTP parsing. This result suggests that trajectory recovery is effective across different vulnerability behaviors, rather than being limited to specific categories.

As illustrated in Table~\ref{tab:CWE}, we compare \appname with \baselinename on the CWEs covered by \baselinename, including path traversal, command injection, code injection, and ReDoS. Prototype Pollution is not applicable because it mainly arises in prototype-based languages such as JavaScript rather than Java. On these predefined CWE types, \appname achieves better performance than \baselinename, outperforming it by 42.9\%, 42.8\%, 50.0\%, and 33.3\%, respectively. This comparison shows that \appname outperforms the SOTA exploit generation method without CWE-specific designs.
We further find two major limitations of \baselinename. 
First, for vulnerabilities that require complex gadget construction, such as XStream deserialization vulnerabilities, \baselinename often fails to construct the required gadget chain. 
Second, \baselinename follows a fixed construction workflow and lacks project-level environment adaptation. As a result, it cannot modify the reproduction environment when the exploit depends on specific runtime conditions, such as switching to a compatible JDK version. In contrast, \appname performs project-level trajectory recovery and refines the exploit toward a valid reproduction.

\vspace{-0.6em}
\begin{center}
\resizebox{\linewidth}{!}{
\begin{tabular}{l!{\vrule width 1pt}p{0.9\columnwidth}}
\makecell{{\large \faLightbulbO}}  &\textbf{Answer to RQ1:} \appname achieves the best overall effectiveness and consistently improves the initial trajectories generated by the base agent. \appname also performs well across diverse CWE categories. \\
\end{tabular}}
\end{center}

\subsection{Ablation Study}

As illustrated in Table~\ref{tab:ablation}, we evaluate \appname from two perspectives: whether the underlying agent can recover ineffective reproductions without guidance, and whether each component of \appname contributes to trajectory recovery.

We first consider two agent-only settings.
\ding{182} \textit{SWE-Agent-Only} directly relies on mini-swe-agent to reproduce the vulnerabilities from the references without any guidance such as differential execution.
\ding{183} \textit{Diff-Feedback-Only} provides the differential execution results to mini-swe-agent when the initial reproduction is finished, and asks it to repair any invalid exploits.
These settings evaluate whether the base agent can either generate exploits directly or use differential feedback by itself to repair ineffective trajectories. We then design three variants to evaluate the contribution of each component to \appname.
\ding{184} \textit{No-Differential-Analysis} uses the vulnerable version execution result for subsequent recovery, without comparing it against the patched version.
\ding{185} \textit{No-Progress-Assessment} analyzes the reproduction process directly without guidance from the three reproduction process dimensions.
\ding{186} \textit{No-Constraint-Guidance} requires the agent to continue the repair from $T_{\text{init}}$ based on the reproduction process without deriving preservation and repair constraints.

\begin{table}[t]
\centering
\caption{Ablation Study Results (Based on Qwen3.5-27B).}
\label{tab:ablation}
\small
\resizebox{\linewidth}{!}{
\begin{tabular}{@{}lcc@{\hspace{0.6em}}|@{\hspace{0.6em}}lcc@{}}
\toprule
\textbf{Variants} & \textbf{Effectiveness} & \textbf{Steps}
& \textbf{Variants} & \textbf{Effectiveness} & \textbf{Steps} \\
\midrule

\appname
& 96 (55.8\%) & 109.3
& w/o Differential
& 69 (40.1\%) & 103.1 \\

Diff-Feedback
& 54 (31.4\%) & 64.3
& w/o Progress
& 85 (49.4\%) & 147.6 \\

SWE-Agent
& 47 (27.3\%) & 47.6
& w/o Constraint
& 72 (41.9\%) & 129.3 \\

\bottomrule
\end{tabular}}
\end{table}

\appname achieves the best performance among all variants, reproducing 96 exploits (55.8\%). We first compare \appname with the agent-only settings. \textit{SWE-Agent-Only} reproduces only 47 exploits (27.3\%), showing that the base agent cannot reliably construct valid exploits from references. Providing differential feedback to the base agent alone only yields a limited improvement, increasing the number of reproduced exploits from 47 (27.3\%) to 54 (31.4\%). This comparison indicates that the base agent cannot effectively interpret differential feedback or guide exploit repairs. Therefore, the improvement of \appname comes from its trajectory analysis and recovery design, rather than the underlying agent or model.

We compare \appname with its ablation variants to understand the contribution of each component. All three variants perform worse than \appname, indicating that differential analysis, progress assessment, and constraint-guided recovery all contribute to the overall effectiveness. The \textit{No-Differential-Analysis} variant achieves only 69 reproductions (40.1\%). Although it uses fewer average steps than \appname, the lower step count mainly results from the absence of patched-version evidence, which causes some simulated exploits to be incorrectly treated as valid reproductions and prevents the agent from further refining ineffective trajectories. The \textit{No-Progress-Assessment} variant achieves 85 reproductions (49.4\%) but requires the most steps. This suggests that progress assessment helps reduce exploratory repairs by decomposing reproduction into predefined dimensions. The \textit{No-Constraint-Guidance} variant drops to 72 reproductions (41.9\%), showing that unfinished tasks alone are insufficient without explicit constraints. 

\vspace{-0.6em}
\begin{center}
\resizebox{\linewidth}{!}{
\begin{tabular}{l!{\vrule width 1pt}p{0.9\columnwidth}}
\makecell{{\large \faLightbulbO}}  &\textbf{Answer to RQ2:} Each component contributes to the effectiveness of \appname, as differential analysis avoids unreliable recovery decisions and progress assessment with constraint guidance keeps recovery focused.  \\
\end{tabular}}
\end{center}

% 1. refploit的表现都超过了每个变体，说明了大家都contribute to了最终的表现

% 2. 每个变体的表现都超过了只使用sweagent，证明了确实是在优化轨迹

% 3. 关于效果与step的平衡，我们发现 SWE only的步数与differnetial变体的步数都少于refploit，swe only是因为没有进行任何修复，而differnetial变体，则因为没有差分，仅通过单侧表现进行是否复现的分析，会容易出现误报，导致没有成功复现的case不会进入recovery loop，因此其步数最低，这个带来的代价是，他在所有的变体中取得了最差的表现

% \subsubsection{Cross-model Generality}

% To evaluate the cross-model generality of \appname, we examine how its performance changes when different stages are replaced with alternative models. As shown in Table~\ref{tab:cross_model_settings}, we consider three settings. \ding{182} In the \textbf{Guidance} setting, the target model is used only for trajectory analysis and guidance generation, while the initial trajectory generation and recovery agent still use the base model. \ding{183} In the \textbf{Takeover} setting, the initial trajectory is generated by the base model, but the target model takes over the subsequent analysis and recovery stages. \ding{184} In the \textbf{Full} setting, the target model is used throughout the entire process. These settings allow us to assess whether the analysis, guidance, and recovery stages of \appname can generalize across different models. %我们在这个阶段选择用qwen作为base model，并通过GPT-5.4-mini以及DeepseekV4来作为替换模型，as？？

\subsection{Cross-agent Generality}

To evaluate whether \appname can be adapted to other code-agent frameworks, we migrate \appname from mini-swe-agent to OpenHands. This adaptation only affects the agent interaction layer, including how \appname segments the agent trajectory, replays a selected trajectory prefix, and launches recovery sub-agents. The core analysis and recovery logic, including differential execution, progress assessment, and constraint generation, remains unchanged.

Under \textit{Qwen3.5-27B}, OpenHands successfully reproduces 51 exploit references before applying \appname. After integrating \appname, the number increases to 89, corresponding to a 74.5\% relative increase. This improvement indicates that \appname can effectively repair ineffective trajectories generated by OpenHands. These results suggest that the trajectory analysis and constraint-guided recovery strategy of \appname can be transferred to another code-agent framework and still bring effective improvement in vulnerability reproduction through trajectory recovery.

\vspace{-0.6em}
\begin{center}
\resizebox{\linewidth}{!}{
\begin{tabular}{l!{\vrule width 1pt}p{0.9\columnwidth}}
\makecell{{\large \faLightbulbO}}  &\textbf{Answer to RQ3:} The improvement on OpenHands shows that \appname can be adapted to other code agents with limited changes. Its trajectory analysis and constraint-guided recovery are not tied to the base agent. \\
\end{tabular}}
\end{center}

% \begin{table}[t]
% \centering
% \caption{Cross-model Evaluation Performance.}
% \label{tab:cross_model_settings}
% \small
% \setlength{\tabcolsep}{3pt}
% \resizebox{\columnwidth}{!}{
% \begin{tabular}{cccc|cccc}
% \toprule
% \textbf{Setting} & \textbf{Initial T.} & \textbf{Guidance} & \textbf{Recovery} & \textbf{GPT-5.4-mini} & \textbf{DeepSeek-V4} \\
% \midrule
% \textbf{Base} & Qwen & Qwen & Qwen & waiting & 47 (27.3\%)  \\
% \textbf{Guidance} & Qwen & Target & Qwen & waiting & waiting \\
% \textbf{Takeover} & Qwen & Target & Target & waiting & waiting   \\
% \textbf{Full} & Target & Target & Target & waiting & 130 (75.6\%) \\
% \bottomrule
% \end{tabular}
% }
% \end{table}
\section{Discussion}

In this section, we qualitatively analyze the recovery behaviors of \appname and discuss the main threats to validity.

\subsection{Qualitative Analysis}

\subsubsection{Successful Recovery Patterns}

To better understand how \appname recovers ineffective agent trajectories, we inspect representative recovery patterns. 

\textbf{Aligning the Harness with Real Attack Surface.}
A particularly important recovery pattern is to redirect a misleading harness to the real attack surface exposed by the library. For example, in CVE-2020-27216, the initial exploit reproduced the race condition through a manually implemented helper that creates the target file rather than invoking the vulnerable temporary-directory creation logic.  \appname repairs it by constructing a real \textit{WebAppContext} and racing to create the temporary directory after Jetty releases the path inside
\textit{WebInfConfiguration.makeTempDirectory()}.  These cases show that \appname succeeds in modifying the harness to the required surface.

\textbf{Ensuring Valid
Payload Semantics.}
  Some initial trajectories fail to reproduce the vulnerability because the
  payload, trigger condition, or validation oracle is semantically misaligned with the reference
  exploit. In such cases, \appname realigns the generated exploit with the expected vulnerability behavior. For example, in CVE-2022-25845, the initial harness used generic Fastjson auto-type payloads \textit{java.util.Properties} and \textit{JdbcRowSetImpl} rather than the required \textit{@type} bypass during the deserialization of the \textit{Exception} or \textit{Throwable} class.
  \appname therefore refines the exploit to the specified bypass payload and validates whether the vulnerable and patched versions diverge under the refined payload. A similar issue appears in CVE-2021-29061, where the initial input fails to trigger the ReDoS
  because it is both too short and structurally inconsistent with the vulnerable regex path, so matching
  never reaches the repeated \textit{(.+:.+@)*} region where the slowdown occurs. \appname repairs this by realigning the URI with the required exploit pattern and
  increasing its length until the vulnerable run reliably times out. These cases show that \appname succeeds in restoring the semantics and structure of the payload to trigger the vulnerability.

\textbf{Repairing Configuration Prerequisites.}
Some initial trajectories fail because settings are missing. For example, in CVE-2022-23457, the initial harness already invoked \textit{ESAPI.validator().getValidDirectoryPath()} with the correct sibling-path bypass input, but the ESAPI initialization failed because the required \textit{ESAPI.properties} configuration is
incomplete. \appname repairs this by supplying a valid ESAPI configuration. A similar issue appears in CVE-2022-25845, where the exploit includes unsupported \textit{--add-opens} flags. \appname repairs this by removing the incompatible runtime options and aligning the environment with the required \textit{fastjson} and JDK combination. These cases show that \appname succeeds in satisfying the prerequisite configuration and runtime environment needed for vulnerability reproduction.

\subsubsection{Failure Case Analysis}
 Not all ineffective trajectories can be recovered. We observe two failure patterns in which \appname substantially improves the trajectory, yet still cannot obtain the expected divergence.

\textbf{Misclassified Progress in Payload Adaptation.} Some failures persist not because the vulnerability path is completely absent, but because the recovered payload is still not fully adapted to the execution environment. In CVE-2021-39144, the recovered payload reaches an XStream deserialization path and raises a \textit{CannotResolveClassException} for \textit{sun.tracing.NullProvider}. This exception should be interpreted as a partial-progress signal during reproduction, suggesting that the payload could be further adapted by adjusting the command used to produce the observable RCE.
However, \appname misclassifies it as an invalid attempt and therefore fails to further adapt the payload.

\textbf{Complex Runtime Configuration Failure.}
  Some failures are dominated by complex runtime prerequisites that consume recovery effort before
  exploit semantics can even be validated. CVE-2021-43113 is a representative example. The exploit
  depends on Ghostscript being visible to the \textit{CompareTool}. Although Ghostscript itself is installed, \textit{ITEXT\_GS\_EXEC} is not propagated correctly, leading to a failed recovery with \textit{``Ghostscript command is not
  specified''}.

\subsection{Threats to Validity}

Our conclusions may be affected by the following threats.

\subsubsection{External validity}
One threat comes from our exclusion of exploits whose key information is only available in images. For example, CVE-2021-40660 contains exploit code in image form. Excluding such cases may limit the generalizability of our results to image-heavy exploit references. We partially reduce this concern by evaluating \appname with \textit{Qwen3.5-27B}, a multimodal-capable model, and confirming that \appname remains effective under this model. However, we have not systematically adapted \appname to process image-only exploit references or evaluated its performance on such cases, and we leave this task to future work.

Although our evaluation uses three large-scale public Java vulnerability datasets, the covered CWEs are associated with only 61.9\% of historical Maven library vulnerabilities. This may limit the generalizability of our results to vulnerability categories that are underrepresented or absent from our datasets. We further find that the uncovered Top CWE categories are mainly associated with authorization issues, such as CWE-862, CWE-284, and CWE-639, or with vulnerabilities that more often arise at the platform or application level rather than the library level, such as CWE-352 and CWE-200. Therefore, the performance of \appname on these vulnerability categories has not yet been evaluated.

\subsubsection{Internal validity}
A major threat comes from our adaptation of \baselinename to Java vulnerabilities. This adaptation may not fully capture all implementation details or design assumptions of \baselinename, and may therefore affect its performance. To mitigate this threat, we preserve the core workflow of \baselinename as much as possible and only modify ecosystem-specific components required for Java reproduction.

Our evaluation environment pre-installs four LTS JDK versions and tools such as \textit{curl}. This setting standardizes the execution environment and allows us to focus on trajectory recovery. However, it also reduces the effort of environment preparation and therefore does not fully evaluate the ability of \appname to install JDKs or prepare tools.

Another internal threat comes from the LLM-based trajectory analysis in \appname. \appname relies on the LLM to judge reproduction progress, map progress judgments to trajectory segments, and generate recovery constraints. Incorrect judgments may lead \appname to preserve misleading trajectory components, which can affect the final recovery outcome. To mitigate this threat, \appname grounds its analysis in differential execution evidence and uses structured progress dimensions to constrain the judgment process.

\section{Related Work}

In this section, we review prior research on software building, vulnerability exploits, and trajectory analysis.

\textbf{Software Building.} Prior studies have extensively investigated the challenges of automatic software building ~\cite{Hassan2018Build,Hassan2017build,Lou2020Building,Yu2025Building,Zhang2022Build,Michael2025Setup}.
Hassan et al.~\cite{Hassan2017build} revealed that nearly half of Java projects fail to build with default commands, while Lou et al.~\cite{Lou2020Building} analyzed Stack Overflow discussions, identifying predominant causes of failure.
Recently, Yu et al.~\cite{Yu2025Building} proposed \textit{CXXCrafter}, an LLM-based agent to resolve build errors for C/C++ open-source software. Bouzenia and Pradel~\cite{Michael2025Setup} proposed \textit{ExecutionAgent}, an LLM-based agent that automatically prepares build scripts across diverse programming languages, build systems, and testing frameworks. However, these works focus on \textit{repairing} existing, well-formed repositories, while disclosed exploits often exist as isolated fragments without a build environment.

\textbf{Library Vulnerability Exploits.}
Library vulnerability exploits are extensively employed in various downstream security tasks.
Existing studies have leveraged exploits to assess vulnerability reachability in client projects~\cite{zirui2024exploiting, Zhou2024Magneto} or to identify affected library versions~\cite{zirui2025poc, Dai2021Exploit}.
However, these works typically rely on manual effort to construct executable exploits.
While recent studies have explored automated one-day exploit generation~\cite{Yang2023oneday,Michael2025PoC, Ray2025PoC}, they are often restricted to specific vulnerability types.
To enhance generality, we explore leveraging LLMs to reproduce vulnerabilities from unstructured disclosure information (e.g., textual advisories).

\textbf{Trajectory Analysis.}
Recent research in agent systems has increasingly focused on understanding failures by analyzing execution trajectories. \textit{MAST}~\cite{cemri2026why} first characterized agent failures across system design, coordination, and task execution, while \textit{TRAIL}~\cite{deshpande2025trailtracereasoningagentic} refined this view with finer-grained categories including reasoning and planning errors. To address these failures, researchers have developed complementary approaches that combine direct inspection, active intervention, and statistical reasoning. \textit{AGDebugger}~\cite{Epperson2025AG} allows developers to steer agent behaviors by editing message histories. \textit{DoVer}~\cite{ma2026dover} actively validates and repairs failures through targeted interventions instead of relying solely on logs. \textit{FAMAS}~\cite{ge2025introducingfailureautomaticallyattributing} replays trajectories and uses spectrum analysis to estimate which agent actions are responsible for failures. However, in the context of vulnerability reproduction, these approaches often fail because execution outcomes are highly diverse and cannot be reliably attributed through single-shot inspection, interventions, or repeated pattern analysis. To address this, we propose a differential analysis approach at both the version and trajectory levels, which evaluates whether a generated intervention is truly effective.

\section{Conclusion}

In this paper, we present \appname, a trajectory recovery framework for facilitating agent-based vulnerability reproduction from publicly disclosed exploit references. \appname analyzes ineffective agent-generated trajectories through differential execution and progress assessment, and derives constraints to guide focused recovery. Evaluated on three open-source Java vulnerability datasets, \appname achieves a reproduction success rate of 80.2\%, outperforming both code-agent baselines and the exploit-generation baseline. Ablation and generality studies further show that its core components contribute to trajectory recovery and that \appname can be applied across different model and agent settings. In future work, we will evaluate \appname with more models and code-agent frameworks to further assess its generality.

\section{Data Availability}
The dataset and source code of \appname will be made publicly available upon acceptance.

\balance
\bibliographystyle{IEEEtran}
\bibliography{main}         

\end{document}